\documentclass[aps,prl,floatfix,twocolumn,amsmath,superscriptaddress]{revtex4-2}
\usepackage{amssymb} 
\usepackage{graphicx}
\usepackage{ulem}
\usepackage{color}
\begin{document}
	
\title{Magnetic-field-induced critical dynamics in magnetoelectric TbPO$_4$}

\author{Ch.\,P. Grams}
\author{M. Grüninger}
\author{J. Hemberger}
	\email[Corresponding author:~]{hemberger@ph2.uni-koeln.de}

\affiliation{
	{\protect II}.\ Physikalisches Institut, Universit\"at zu K\"oln, Z\"ulpicher Str.\ 77, D-50937 K\"oln, Germany}

\begin{abstract}
We report on the magnetoelectric dynamics in the linear magnetoelectric antiferromagnet TbPO$_4$ studied by broadband dielectric spectroscopy. 
For the phase transition into the magnetoelectric antiferromagnetic phase at $T_{N} \approx 2.3$\,K, a finite magnetic field $H$ induces critical behavior in the quasi-static permittivity $\varepsilon^\prime(T)$. Plotting the corresponding anomaly as function of $T/T_N(H)$, 
we observe the scaling behavior $\Delta \varepsilon^\prime \propto H^2$, a clear fingerprint of linear magnetoelectric antiferromagnets. Above the phase transition, we find
a critical slowing down of the ferroic fluctuations in finite magnetic field. This behaviour can be understood via a magnetic-field-induced relaxational response that  
resembles the soft-mode behaviour in canonical ferroelectrics and multiferroics. 
\end{abstract}

\date{December 1, 2023}

\maketitle

Magnetoelectric multiferroics exhibit coupled magnetic and electric order, and the coupling manifests in the occurrence of a common order parameter \cite{Khomskii2009,Spaldin2005,Tokura2014,Dong2015,Fiebig2016,Spaldin2019}. 
Asides the question on coupling mechanisms leading to such complex ordering phenomena, the dynamics of corresponding fluctuations and the critical behavior of coupled ferroic order parameters \cite{Kim2014} are interesting aspects of investigation. One kind of elementary excitations within a magnetoelectric multiferroic phase are electromagnons which combine the spin-wave excitation of the magnetic structure with an electric-dipole active contribution such as a polar lattice distortion \cite{Mochizuki2010}.  
Such electromagnons may be associated with the symmetry-breaking Goldstone modes of magnetoelectric multiferroic order \cite{Katsura2007,Pimenov2006}.  
This implies that such modes can be expected to soften near a second-order phase transition, analogous to the case of canonical ferroelectric materials \cite{Blinc1974-, Kamba2021}. Therefore, these soft modes should influence the dynamical response of fluctuations above the onset of static multiferroic order \cite{Shuvaev2010}.

Critical slowing down of magnetoelectric fluctuations has been reported near a multiferroic phase transition \cite{Niermann2015} and even near a multiferroic quantum phase transition \cite{Grams2022}.
The latter raises the question whether long-range magnetoelectric multiferroic order is required for magnetoelectric fluctuations exhibiting critical slowing down.
We address this question in the linear magnetoelectric antiferromagnet TbPO$_4$, in which antiferromagnetic (AFM) order is \textit{not} accompanied by a spontaneous electric polarization $P$ in zero magnetic field. 
Instead, the polarization $P$ exhibits a contribution proportional to the applied magnetic field $H$ according to the linear magnetoelectric effect. This implies that the magnetic field alters the magnetic structure in a way that breaks spatial inversion symmetry \cite{Rivera2009}. 
In this article, we examine the fluctuation dynamics of this field-induced polarization in the vicinity of the AFM phase transition. 
In particular, we report on spectroscopic investigations of the complex permittivity $\varepsilon^*(H,T,\nu)$ 
in high-quality single-crystalline TbPO$_4$ for frequencies up to 3\,GHz in order to shed light on the dynamical dielectric response of the magnetoelectric fluctuations induced by external magnetic fields.

\begin{figure}[b]
	\centering
 \includegraphics[width=0.45\columnwidth,angle=0]{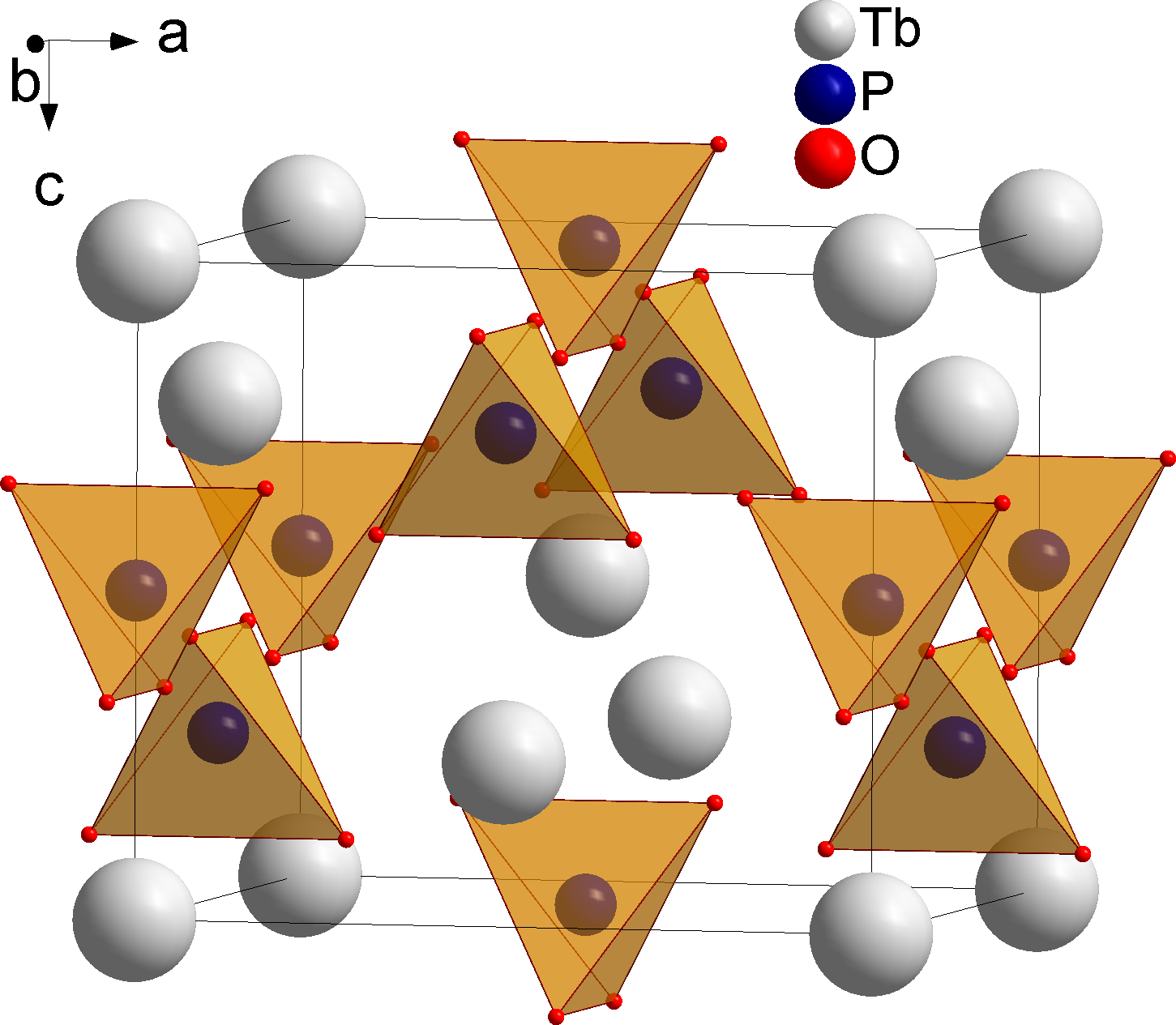}
 \hfill
 \includegraphics[width=0.52\columnwidth,angle=0]{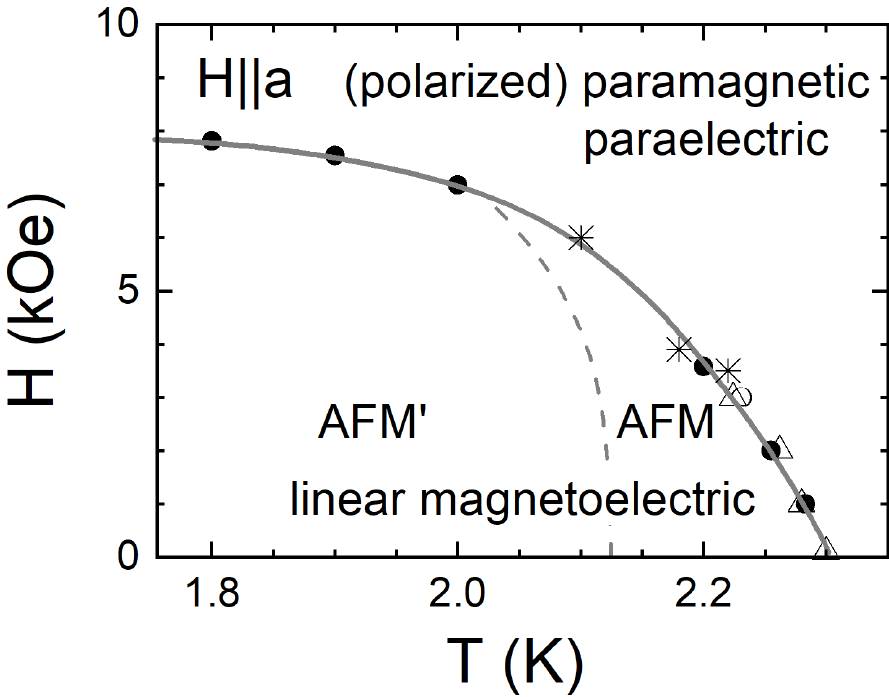}
	\caption{
	  Left: Sketch of the tetragonal unit cell of TbPO$_4$.
    Right: $(H,T)$ phase diagram for the linear magnetoelectric antiferromagnetic phases in TbPO$_4$. 
    The symbols were derived from measurements (not shown) of $P(H,T)$ ($\bullet$), $\varepsilon^*(H,T)$ ($\triangle$), and $\varepsilon^*(H,T,\nu)$ ($\ast$). The lines visualize literature results  gained by neutron scattering, birefringence, magnetoelectric, and magnetic susceptibility measurements \cite{Mensinger1993,Muller1993,Becker1985-}. 
	}
	\label{fig_phase}
\end{figure}

TbPO$_4$  belongs to the above described class of linear magnetoelectric antiferromagnets \cite{Rivera2009}
in which electric polarization is absent for $H$\,=\,0 but can be induced according to 
$P$\,=\,$\alpha\, H$ within the antiferromagnetically ordered phase. 
In fact, TbPO$_4$ is the current record holder among single-phase multiferroics with a magnetoelectric coefficient 
$|\alpha| \approx 730$\,ps/m at 1.5\,K \cite{Ying2022,Rivera2009}, 
making it an excellent model system for our study. 
The system crystallizes in the tetragonal zircon-type structure with space group $I4_1/amd$ and point symmetry $\bar{4}m2$ for the Tb$^{3+}$ sites \cite{Bluck1988} and can be described as staggered chains of alternating Tb$^{3+}$ ions and tetrahedral PO$_4^{3-}$ entities along the $c$ axis, see Fig.\ \ref{fig_phase}. 
The spin system undergoes a sequence of phase transitions, 
and the corresponding $(H,T)$ phase diagram is sketched in Fig.\,\ref{fig_phase}.
Cooling down from the paramagnetic phase in $H$\,=\,0, 
antiferromagnetic order with the spin axis along $c$ is established at $T_{N}$\,=\,2.28\,K.\@ 
At $T_{N'}\approx 2.13$\,K the spin axis is tilted away from [001] and the system looses tetragonal symmetry but still keeps its collinear AFM character \cite{Naegele1980}. 
Finite $H$ leads to spin canting which via local exchange striction causes the onset of electric polarization.
Above a direction-dependent critical field $H_c$ the spins turn into the field direction and a spin-polarized paramagnetic phase is reached. 
Both AFM phases show the linear magnetoelectric effect \cite{Rivera2009}
revealing slightly different magnetoelectric coefficients depending on the field direction. 
In our measurements of the complex dielectric constant  $\varepsilon^*(H,T)$ or of $P(H,T)$ the transition between the two phases is hardly visible. 
We will focus on magnetoelectric fluctuations above the transition into the AFM phase.

\begin{figure}[t]
	\centerline{\includegraphics[width=0.999\columnwidth,angle=0]{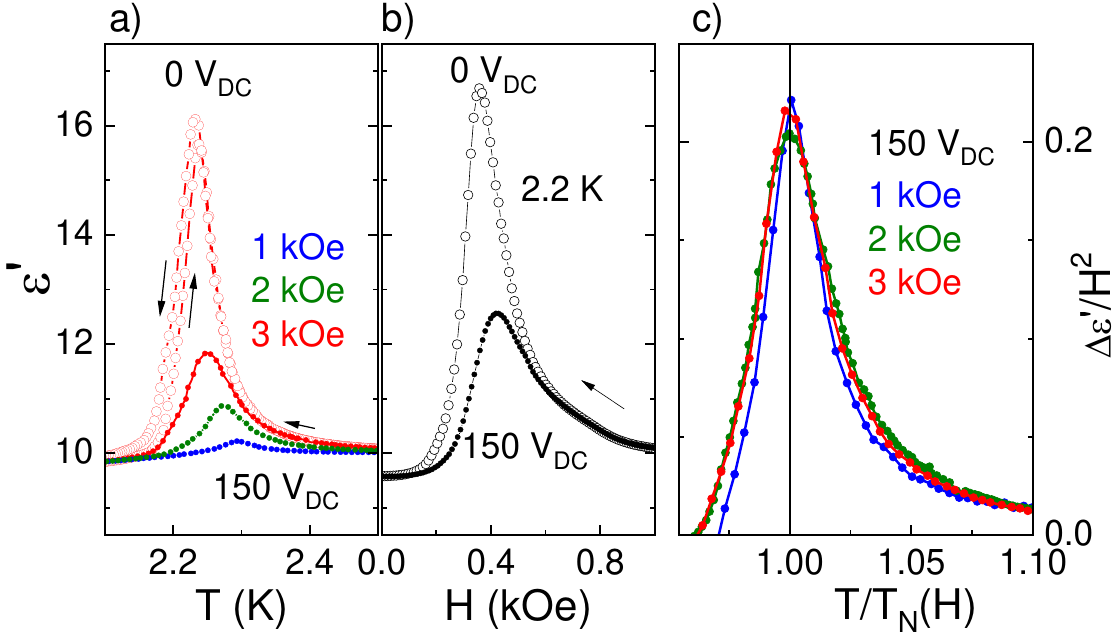}}
	\caption{
    Critical behavior in the quasi-static permittivity $\varepsilon^\prime(T,H)$ at $\nu$\,=\,113\,Hz. 
    a) Data of $\varepsilon^\prime(T)$ measured with zero electric bias in $H$\,=\,3\,kOe ($\circ$) show diverging character. Curves with a dc bias of 150\,V$_{\rm DC}$ equivalent to $E\approx 540$\,V/mm ($\bullet$) are plotted for various $H$ fields. 
	b) $\varepsilon^\prime(H)$ at $T$\,=\,2.2\,K with and without electric bias. 
    c) Scaling plot of the same data (after subtracting $\varepsilon_\infty\approx10$) normalized to $H^2$, plotted as function of $T/T_N(H)$.    
	}
	\label{fig_LFeps}
\end{figure}

The single crystal of TbPO$_4$ used for this study was flux grown and has been characterized in \cite{Rivera2009}. Details of the growth procedure are described elsewhere \cite{Hintzmann1969}. 
The sample was prepared as small platelet in capacitor geometry with dimensions of thickness $d\!\approx \!280$\,$\mu$m and cross section $A\!\approx \!1.45$\,mm$^2$ oriented in the (100) plane. 
The electrodes were formed by evaporating gold. 
The dielectric measurements were made in a commercial $^4$He flow cryo-magnet ({\sc Quantum-Design PPMS}) employing a home-made 50\,$\Omega$ coaxial-line inset with a sample holder in coplanar waveguide geometry.
For frequencies  100~kHz~$\leq\nu\leq$~3~GHz we used a vector network analyzer ({\sc ZNB8, Rohde\&Schwarz}) and evaluated the complex impedance $Z^*(\nu)$ via the scattering coefficients $S_{12}$ and $S_{11}$. 
The complex, frequency-dependent permittivity 
$\varepsilon^*(\nu)$\,=\,$\varepsilon^\prime + i \varepsilon^{\prime\prime}$ is related to $Z^*(\nu)$\,=\,$1/\left( i 2 \pi \nu C_0 \varepsilon^*(\nu)\right) $ with the geometric capacitance 
$C_0$\,=\,$\varepsilon_0 A/d$ given by the sample geometry. 
The effectively applied ac voltage in this setup was 0.22\,V$_{rms}$. 
For lower frequencies we used a high-impedance frequency response analyzer [{\sc Novocontrol}] together with respective high-voltage amplifier modules in order to apply dc bias voltages or higher ac stimuli up to 200\,V.\@
In the tetragonal AFM phase, the non-vanishing elements of the magnetoelectric tensor $\alpha_{ij}$ in $P_i$\,=\,$\alpha_{ij} H_j$ 
are $\alpha_{aa}$\,=\,$-\alpha_{bb}$ \cite{Rado1984,Bluck1988}. 
All  measurements were performed with electric and magnetic fields along the crystallographic $a$ axis. 
We found a value of $\alpha_{aa}\approx 280$~ps/m at $T=2.2$\,K.

In ferroelectric or multiferroic compounds with spontaneous electric polarization one expects a 'diverging' quasi-static dielectric permittivity at the corresponding second-order phase transition.  
Actually, even in ferroelectrics $\varepsilon^\prime(T)$ will stay finite at $T_c$ due to damping \cite{Blinc2011-}. 
In type-II multiferroics in which the ferroelectric component reflects only a secondary order parameter, the anomaly at the critical temperature often turns out to be smaller than the dielectric background 
$\varepsilon_\infty$ \cite{Tokura2010,Schrettle2009}. 
In linear magnetoelectric antiferromagnets, electric and magnetic fields are linearly coupled to the free energy $F\propto L \cdot E\,H$ via the primary antiferromagnetic order parameter $L$. 
This results in $P=\partial F/\partial E \propto L \cdot H$ 
and $M=\partial F/\partial H\propto L \cdot E$ 
within the magnetoelectric AFM phase.
For $H$\,=\,0, neither spontaneous polarization nor a contribution to $\varepsilon^\prime$ are expected, assuming that no lattice distortion is involved in the AFM transition.
However, in finite magnetic field the onset of finite polarization at $T_N$ and thus critical behavior of $\varepsilon^\prime$ can be expected.
Figure \ref{fig_LFeps}a) shows quasi-static, low-frequency data of $\varepsilon^\prime(T)$ at $\nu$\,=\,113\,Hz in different magnetic fields. 
The electrically unbiased data, measured in a magnetic field of, e.g, $H$\,=\,3\,kOe (red $\circ$), exhibit a sharp peak at the AFM transition, i.e., 'divergent' behavior.
Below $T_N$ a small difference between cooling and heating 
is observed which can be attributed to domains.
Applying an additional electric bias field 
stabilizes the magnetoelectric phase that exhibits finite electric polarization.  
This stabilization shifts the peak position -- the onset of AFM order -- in $\varepsilon^\prime(T)$ to higher $T$ and in $\varepsilon(H)$ to larger $H$, see Fig.\,\ref{fig_LFeps}b). Furthermore, the dc bias voltage flattens this peak as polarization is finite already above the AFM transition.

We find that the strength of the magnetoelectric anomaly in $\varepsilon^\prime$ depends on the applied magnetic field, scaling with $H^2$ near the AFM transition. For the corresponding scaling plot in Fig.\,\ref{fig_LFeps}c), we employ 
$\Delta\varepsilon^\prime$\,=\,$\varepsilon^\prime-\varepsilon_\infty$ as a measure of critical fluctuations. In the GHz range, the constant term $\varepsilon_\infty \approx 10$ denotes the polarizability of phonons and of electronic excitations.  Using the field-dependent transition temperature $T_N(H)$ and plotting $\Delta \varepsilon^\prime / H^2$ vs.\ $T/T_N(H)$, the data collapse onto a single curve, see Fig.\,\ref{fig_LFeps}c). 
This behavior has been explained by Mufti {\it et al.}\ \cite{Mufti2011} using a free-energy expansion to describe the influence of spin fluctuations on the quasi-static dielectric response in antiferromagnets showing the linear magnetoelectric effect. In turn, the observed scaling behavior is a clear fingerprint of linear magnetoelectric antiferromagnets \cite{Mufti2011} and highlights the presence and strength of critical spin fluctuations above the linear magnetoelectric AFM phase.

\begin{figure}[t]
	\centerline{\includegraphics[width=0.999\columnwidth,angle=0]{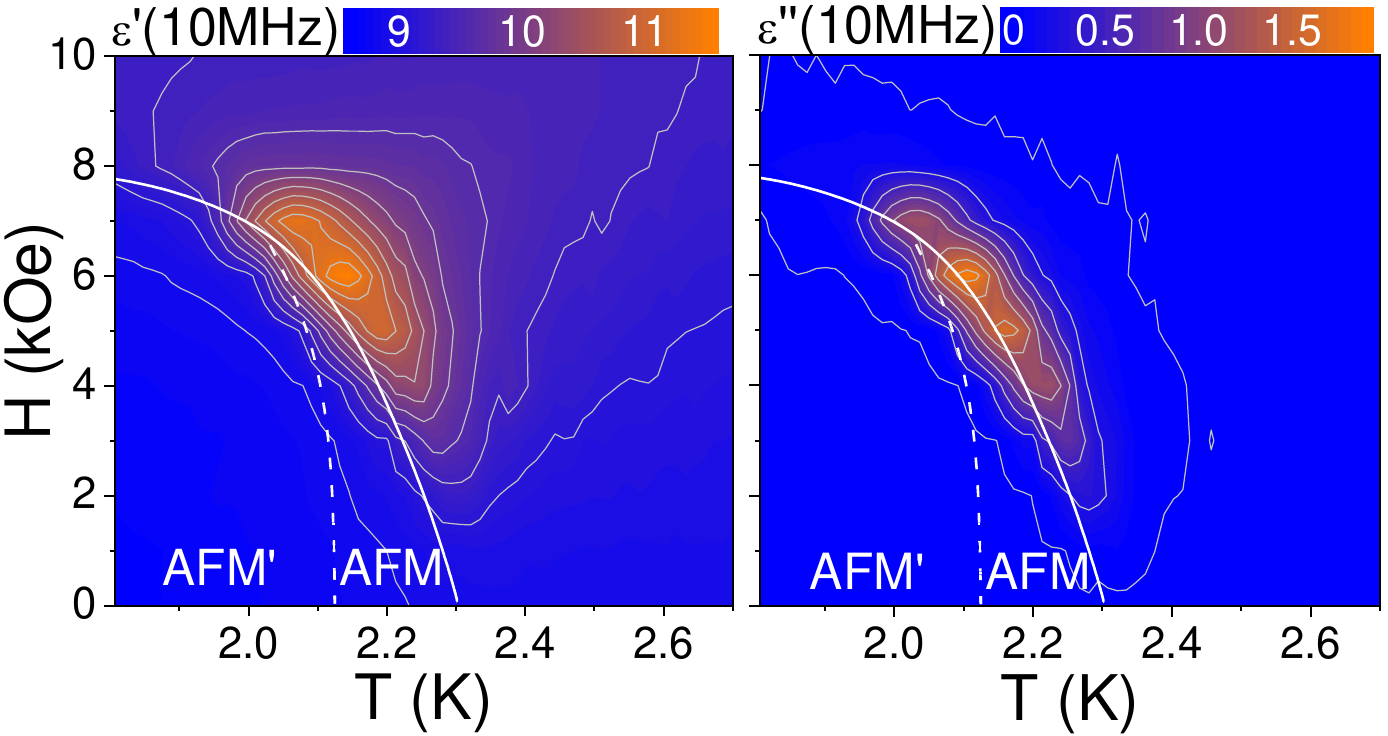}}
	\caption{
		Contour plots of the complex permittivity $\varepsilon^*(T,H)$ around the AFM phase transition, measured at $\nu$\,=\,10\,MHz (left/right: real/imaginary part). 
	}
	\label{fig_contour}
	\vspace{2mm}
 \centerline{\includegraphics[width=0.999\columnwidth,angle=0]{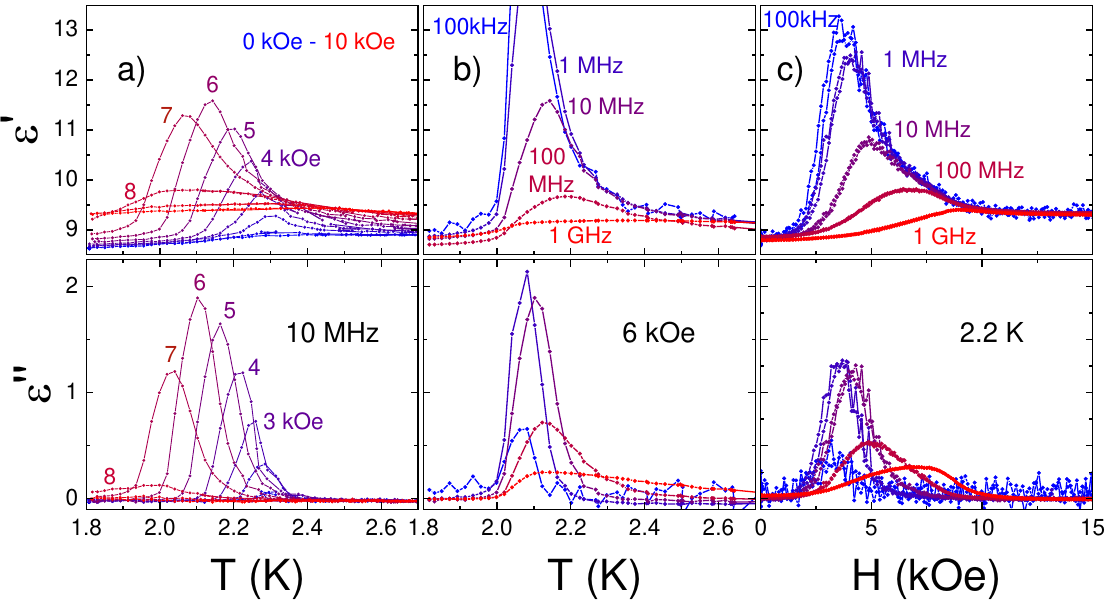}}
	\caption{
    One-dimensional cuts through $\varepsilon^\prime(\nu,H,T)$ (top) and     $\varepsilon^{\prime\prime}(\nu,H,T)$ (bottom), compare Fig.\ \ref{fig_contour}.
    a) $\varepsilon^*(10\,{\rm MHz},T)$,  
    b) $\varepsilon^*(6\,{\rm kOe},T)$, 
    and c) $\varepsilon^*(H, 2.2\,{\rm K})$. 
    No electric bias field has been used.}
	\label{fig_epsfreq}
\end{figure}

We focus on the dynamics of these critical fluctuations in the vicinity of the AFM phase boundary and therefore turn to spectroscopic measurements at higher frequencies. 
We address data of the complex dielectric permittivity $\varepsilon^*(\nu,H,T)$ 
measured without an electric bias field. These data, measured in a three-dimensional parameter space, are illustrated in Fig.\ \ref{fig_contour} via two-dimensional plots of 
$\varepsilon^\prime(T,H)$ and $\varepsilon^{\prime\prime}(T,H)$ for fixed frequency $\nu$. In these plots, the dielectric response nicely retraces the $(H,T)$ phase diagram given in Fig.\,\ref{fig_phase}. Above the transition to the AFM phase, the strength of the response increases with $H$, as discussed above for low frequencies. 
To better demonstrate the quantitative behavior, we plot one-dimensional cuts in Fig.\,\ref{fig_epsfreq}, showing $\varepsilon^*(T)$ for different $H$ in panel a) and for different $\nu$ in b) as well as $\varepsilon^*(H)$ in c). 
The loss peaks in $\varepsilon^{\prime\prime}$ are located 
just above the phase transition where they meet the low-$T$ falling slope of the feature in the real part $\varepsilon^\prime$.

For low frequencies such as 100\,kHz, $\varepsilon^\prime(T)$ shows the critical, i.e., divergent behavior expected for a second-order phase transition, 
see Fig.\ \ref{fig_epsfreq}b). This reflects the diverging spatial correlation length. 
To describe the dynamical, frequency-dependent response, the behavior of the correlation time of the fluctuations has to be considered as well. 
In the loss $\varepsilon^{\prime\prime}$, a relaxation peak appears above $T_N$ when the experimental frequency of the stimulus meets the fluctuation rate. With increasing temperature, the maximum loss occurs for higher frequencies, see Fig.\ \ref{fig_epsfreq}b), indicating the decrease of the correlation time. The same behavior is observed with increasing $H$ in Fig.\ \ref{fig_epsfreq}c).

\begin{figure}
	\centerline{\includegraphics[width=0.999\columnwidth,angle=0]{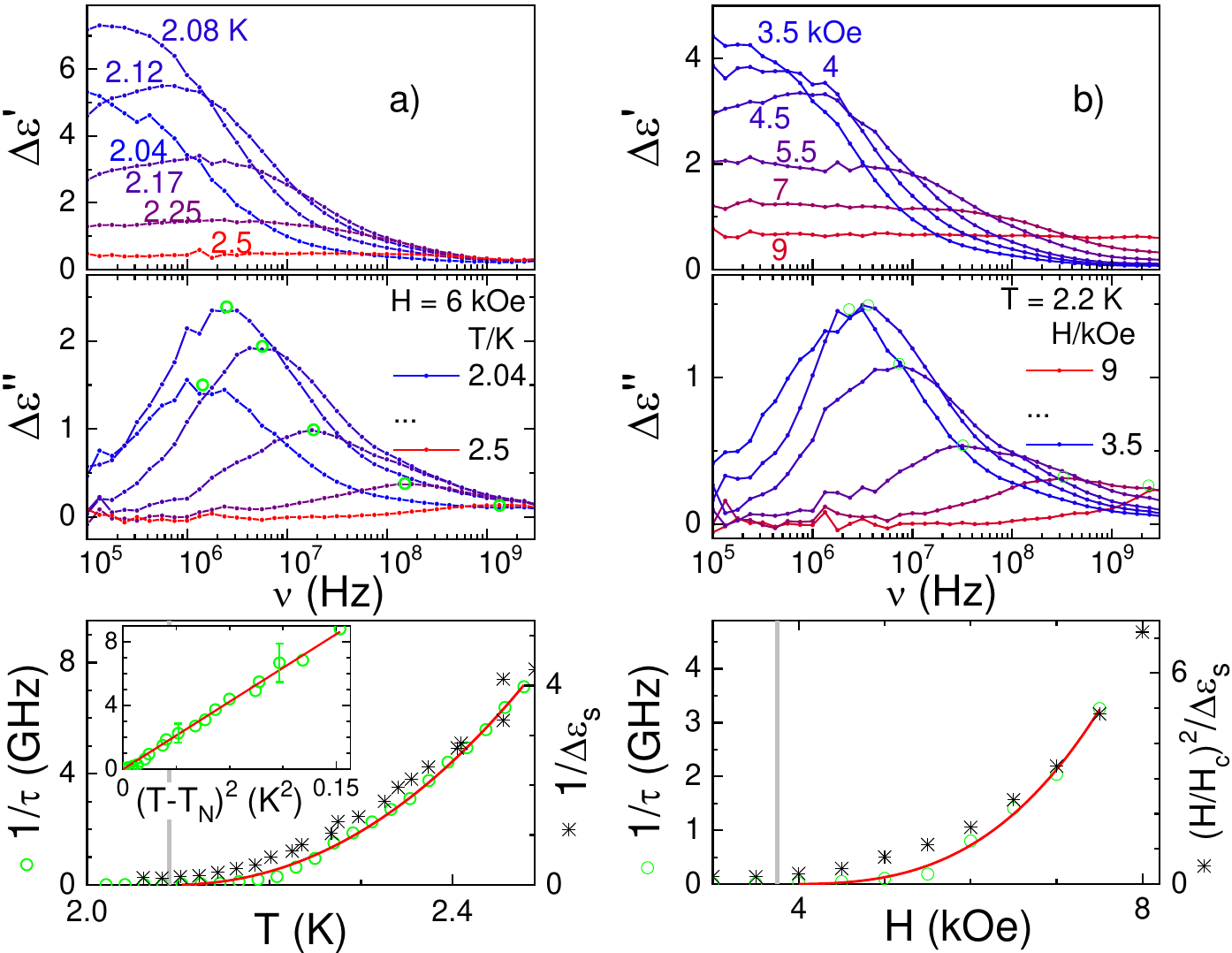}}
	\caption{
	Spectra of the complex permittivity $\varepsilon^*(\nu)$ for constant $T$ and $H$. 
    Top/middle panels: real/imaginary part.
    a) Result for $H$\,=\,6\,kOe for several temperatures. From each curve, the data for $H$\,=\,6\,kOe and $T$\,=\,3\,K have been subtracted.
    b) Result for $T$\,=\,2.2\,K for several magnetic fields $H$. The data for 2.2\,K and $H$\,=\,0 have been subtracted. 
    Bottom panels: Mean relaxation rate $1/\tau_c$ determined from the peak positions in $\varepsilon^{\prime\prime}$ (green symbols). 
    For comparison, we plot $1/\Delta\varepsilon_s$ for the $T$ dependence in the left panel or the normalized value $(H/H_c)^2/\Delta\varepsilon_s$ for the field dependence in the right panel, respectively.   
    Grey vertical lines mark the transition temperature $T_N$($H$\,=\,6\,kOe) and critical field $H_c$($T$\,=\,2.2\,K), respectively. 
    Inset: $1/\tau_c$ vs.\ $(T-T_N)^2$.
    }
	\label{fig_spectra}
\end{figure}

For a quantitative discussion of this dispersive behavior, it is most appropriate to consider the frequency dependence for constant $T$ and $H$ in order to separate, e.g., the field dependence and temperature dependence of the strength of the response. Such spectra of the complex permittivity $\varepsilon^*(\nu)$ are shown in Fig.\ \ref{fig_spectra}. 
The data in the left panels, Fig.\ \ref{fig_spectra}a), is taken at constant $H$\,=\,6\,kOe for different temperatures. To focus on the magnetoelectric contribution, we subtracted the data measured at 
$T$\,=\,3\,K and $H$\,=\,6\,kOe. We observe a step in $\varepsilon^\prime(\nu)$ that is accompanied by a loss peak in $\varepsilon^{\prime\prime}$. 
This behavior can be described as Debye relaxation, i.e., 
in terms of an overdamped harmonic oscillator \cite{Blinc1974-,Niermann2015}, 
\begin{equation}
    \varepsilon^\prime  =  \varepsilon_\infty + 
    \frac{\Delta \varepsilon_s}{1+ \omega^2\tau_c^2} \, ,
    \hspace{8mm}
    \varepsilon^{\prime\prime}  =   
    \frac{\Delta \varepsilon_s\, \omega \tau_c}{1+ \omega^2\tau_c^2} 
    \label{eq:Debye}
\end{equation}
with $\omega$\,=\,$2\pi \nu$ and the step height or relaxation strength $\Delta\varepsilon_s$ 
with $\varepsilon^\prime(\omega\!=\!0)$\,=\,$\varepsilon_\infty +\Delta\varepsilon_s$. 
The effective relaxation time $\tau_c$ represents the fluctuation lifetime and 
increases on approaching the continuous, second-order phase transition.  
The relaxation rate $1/\tau_c$ can be read off directly from the peak in $\varepsilon^{\prime\prime}$ which is located at $\omega$\,=\,$1/\tau_c$.

Upon cooling towards $T_N(H)$ in constant, finite magnetic field $H$, the step height or relaxation strength $\Delta\varepsilon_s$ observed in the spectra increases according to the divergent behavior of the quasi-static permittivity. Below $T_N(H)$ the step height decreases again, 
but our focus is on the critical behavior above the phase transition. 
We determine the temperature dependence of $1/\tau_c$ from the peaks in $\varepsilon^{\prime\prime}$, see green symbols in Fig.\,\ref{fig_spectra}, and the result is shown in the bottom panel. 
With decreasing temperature, the relaxation rate continuously diminishes approaching the AFM transition. 
This result clearly establishes the critical slowing down of magnetoelectric fluctuations in a linear magnetoelectric antiferromagnet.

Obviously, this critical slowing down of the fluctuation dynamics is related to the critical increase of the quasi-static permittivity towards the phase transition. 
In fact, $1/\tau_c$ and $1/\Delta\varepsilon_s$ exhibt the same temperature dependence, as shown in the bottom panel of Fig.\ \ref{fig_spectra}.
This can be understood in analogy to the softening of polar lattice modes in proper ferroelectrics, 
which is described by the Lyddane-Sachs-Teller relation $\Delta\varepsilon_s \propto \omega_0^{-2}$ connecting the angular eigenfrequency of the undamped mode $\omega_0$ with its oscillator strength
$\Delta\varepsilon_s$ \cite{Blinc1974-}. In the overdamped case, where the damping $\Gamma$ is comparable to or larger than $\omega_0$, the resonant character of the excitation turns into 
the relaxation behavior described by Eq.\ (\ref{eq:Debye}) with the critical timescale $\tau_c\approx \Gamma/\omega_0^2$. 
This overdamped scenario is appropriate when, e.g., $\omega_0$ is the eigenfrequency of 
a mode that softens in the vicinity of a phase transition. 
For relaxation behavior, the Lyddane-Sachs-Teller relation transforms to $\varepsilon_s \propto \tau_c$, which corresponds to the mean field result for dynamic critical scaling \cite{Hohenberg1977} 
and agrees with our experimental result.

Critical slowing down is described by $1/\tau_c\propto(T-T_{N})^{\gamma}$. 
In the present case the data can be fitted using a critical exponent of $\gamma = 2.1 \pm 0.4$, 
see inset of bottom panel of Fig.\ \ref{fig_spectra}. The dominant contribution to the sizable error bar stems from the uncertainty of the precise value of $T_N$. Despite the error bar it can be stated that this value is larger than the canonical expectation of $\gamma$\,=\,1 for proper ferroelectrics \cite{Blinc1974-}. This is not unusual for magnetic materials where values larger than unity 
are expected. In chiral multiferroics like MnWO$_4$ or LiCuVO$_4$ values of $\gamma\approx 1.3$ have been found \cite{Niermann2015,Grams2022}. In TbPO$_4$, however, $\gamma$ 
appears to be even higher. This may point to a stronger influence of the quantum nature of the critical fluctuations studied here, as the critical temperatures realized in TbPO$_4$ are considerably lower than 
in the multiferroic examples mentioned above.
A value of $\gamma\approx 2$ also was found near multiferroic quantum phase transitions \cite{Grams2022,Kim2014}.

Critical slowing down of the magnetoelectric fluctuation 
can also be observed on approaching the AFM phase boundary at constant temperature by decreasing the external magnetic field, as illustrated in Fig.\,\ref{fig_spectra}b). 
The general picture is very similar to the temperature-driven scenario just discussed. Starting from above the critical field 
$H_c(T\!=\!2.2\,{\rm K})$\,=\,3750\,Oe, a step-like contribution to $\varepsilon^\prime(\nu)$ evolves on lowering the magnetic field, accompanied by a peak in the dielectric loss $\varepsilon^{\prime\prime}(\nu)$. The peak position in $\varepsilon^{\prime\prime}(\nu)$ 
shifts more and more towards lower frequencies on approaching the AFM transition, denoting the slowing down of the magnetoelectric fluctuations. 
Analogous to the case of temperature as control parameter, 
we extract the critical fluctuation rate $1/\tau_c(H)$ from the loss maxima. 
A quantitative analysis of the field dependence, however, has to cope with possible effects of the demagnetization factor, in particular for a plate-like sample. We hence refrain from determining a critical exponent. Note that this does not affect the qualitative picture of critical slowing down.
To connect this critical behavior of the correlation time $\tau_c$ to the quasi-static behavior, which is determined by the correlation length of the magnetoelectric fluctuations, one has to consider the magnetic-field dependence of the magnetoelectric contribution. As shown above, the relaxation strength $\Delta\varepsilon_s(H)$ is proportional to $H^2$. Therefore, we normalized the quasi-static contribution $\Delta\varepsilon_s$  by $H^2$ as depicted on the right scale of the bottom panel of Fig.\,\ref{fig_spectra}. 
This yields reasonable but not perfect agreement with the result for $1/\tau_c$.

Summarizing, we revealed the critical dynamics of magnetoelectric fluctuations in a linear magnetoelectric antiferromagnet via broadband dielectric spectroscopy up to GHz frequencies.
Above the continuous phase transition into long-range antiferromagnetic order in TbPO$_4$, a slowing down of the magnetoelectric fluctuations can be monitored in finite external magnetic field via the evaluation of relaxational contributions to the complex permittivity.  
Above $T_{N}$, the corresponding loss spectra $\varepsilon''(\nu)$ show characteristic maxima from which the fluctuation rate $1/\tau_c$ can be determined. 
Upon approaching $T_N(H)$ by lowering the temperature in finite magnetic field, the relaxation rate vanishes and the fluctuation lifetime diverges. 
The data can be described with a critical exponent $\gamma$\,=\,$2.1\pm 0.4$,
which is larger than in multiferroic systems. 
Furthermore, we have shown that slowing down of the magnetoelectric fluctuations can not only be driven by temperature but also via an external magnetic field.

The quasi-static relaxation strength $\Delta\varepsilon_s$ gives a 'divergent' contribution to the permittivity which scales with the fluctuation lifetime $\tau$.
This scenario can be understood in analogy to the softening of an overdamped polar mode in ferroelectrics, in which the Lyddane-Sachs-Teller relation couples the dielectric oscillator strength to the effective relaxation rate. 
However, in the present case of a linear magnetoelectric antiferromagnet the underlying fluctuations are of predominantly magnetic origin and thus the relaxation strength scales with the square of the magnetic field, $\Delta\varepsilon_s\propto H^2$.
It will be interesting to compare these dynamical characteristics of the linear magnetoelectric antiferromagnet TbPO$_4$ to the corresponding behavior of other magnetoelectrics possessing different coupling mechanisms.

\begin{acknowledgements}
The authors thank J.-P. Rivera (University of Geneva) for the provision of the well-characterized sample and helpful discussions.
Furthermore, we acknowledge funding from the Deutsche Forschungsgemeinschaft (DFG, German Research Foundation) through Project No.\ 277146847 -- CRC 1238 (project B02). 
\end{acknowledgements}

\bibliography{TbPO4.bib} 

\end{document}